\documentstyle [12pt] {article}
\title {A SIMPLE, APPROXIMATE METHOD FOR ANALYSIS OF KERR-NEWMAN BLACK HOLE DYNAMICS AND THERMODYNAMICS}

\author{Vladan Pankovi\'c$^{\ast,\sharp}$,
Simo Ciganovi\'c$^\sharp$, Rade Glavatovi\'c $^\diamond$ \\
$^\ast$Department of Physics, Faculty of Sciences, 21000 Novi
Sad,\\ Trg Dositeja Obradovi\'ca 4. , Serbia, vdpan@neobee.net \\
$^\sharp$Gimnazija, 22320 Indjija, Trg Slobode 2a, Serbia \\
$^\diamond$Military-Medical Academy, 11000 Belgrade, Crnotravska
17., Serbia}

\date {}
\begin{document}
\maketitle

\vspace {0.2cm} PACS number: 04.70.Dy \vspace {0.2cm}

\begin {abstract}
In this work we present a simple, approximate method for analysis
of the basic dynamical and thermodynamical characteristics of
Kerr-Newman black hole. Instead of the complete dynamics of the
black hole self-interaction we consider only such stable
(stationary) dynamical situations determined by condition that
black hole (outer) horizon circumference holds the integer number
of the reduced Compton wave lengths corresponding to mass spectrum
of a small quantum system (representing quant of the black hole
self-interaction).  Then, we show that Kerr-Newman black hole
entropy represents simply the quotient of the sum of static part
and rotation part of mass of black hole on the one hand and ground
mass of small quantum system on the other hand. Also we show that
Kerr-Newman black hole temperature represents the negative value
of the classical potential energy of gravitational interaction
between a part of black hole with reduced mass and small quantum
system in the ground mass quantum state. Finally, we suggest a
bosonic great canonical distribution of the statistical ensemble
of given small quantum systems in the thermodynamical equilibrium
with (macroscopic) black hole as thermal reservoir. We suggest
that, practically, only ground mass quantum state is significantly
degenerate while all other, excited mass quantum states are
non-degenerate. Kerr-Newman black hole entropy is practically
equivalent to the ground mass quantum state degeneration. Given
statistical distribution admits a rough (qualitative) but simple
modeling of Hawking radiation of the black hole too.

\end {abstract}

\section {Introduction}

In this work, generalizing our previous result on Schwarzschild
and Kerr-Newman black hole [1]-[3], we shall present a simple,
approximate method for analysis of the basic dynamical and
thermodynamical characteristics (Bekenstein-Hawking entropy and
Hawking temperature) of  Kerr-Newman black hole. Instead of the
complete dynamics of Kerr-Newman black hole self-interaction we
shall consider only such stable (stationary) dynamical situations
determined by condition that Kerr-Newman black hole (outer)
horizon circumference holds the integer number of the reduced
Compton wave lengths corresponding to mass spectrum of a small
quantum system (representing quant of Kerr-Newman black hole
self-interaction). (Obviously it is conceptually analogous to Bohr
quantization postulate interpreted by de Broglie relation in Old,
Bohr-Sommerfeld, quantum theory. Also, it can be pointed out that
our formalism is not theoretically dubious, since, at it is not
hard to see, it can represent an extreme simplification of a more
accurate, e.g. Copeland-Lahiri [4], string formalism for the black
hole description.) Then, we shall show that Kerr-Newman black hole
entropy represents the quotient of the sum of static
(Schwarzschild) part and rotation part of mass of Kerr-Newman
black hole on the one hand and ground mass of small quantum system
on the other hand. Also we shall show that black hole temperature
represents the negative value of the classical potential energy of
gravitational interaction between a part of black hole with
reduced mass and small quantum system in the ground mass quantum
state. Finally, we shall suggest a bosonic great canonical
distribution of the statistical ensemble of given small quantum
systems in the thermodynamical equilibrium with (macroscopic)
Kerr-Newman black hole as thermal reservoir. We shall suggest
that, practically, only ground mass quantum state is significantly
degenerate while all other, excited mass quantum states are
non-degenerate. Kerr-Newman black hole entropy is practically
equivalent to the ground mass quantum state degeneration. Given
statistical distribution admits a rough (qualitative) but simple
modeling of Hawking radiation of the black hole too. In many
aspects this modeling is very close to Parikh and Wilczek modeling
of Hawking radiation as tunneling [5].

\section {Theory}

As it is well-known [6] outer horizon radius of Kerr-Newman black
hole is given by expression
\begin {equation}
 R = M + (M^{2} - {\it a}^{2}- Q^{2})^{\frac {1}{2}}
\end {equation}
where $M$ represents the black hole mass, ${\it a}= \frac {J}{M}$
where $J$ represents the black hole angular momentum, while $Q$
represents the black hole electric charge. It implies
\begin {equation}
 M = \frac {R}{2}+ \frac {1}{2}\frac {{\it a}^{2}}{R} + \frac {1}{2}\frac {Q^{2}}{R} = M_{s} + M_{r} + M_{c}        .
\end {equation}

First part of $M$, $M_{s} = \frac {R}{2}$, can be considered as an
effective black hole mass corresponding to a fictitious
Schwarzschild black hole with horizon radius $R$. In fact $M_{s}$
can be considered as the mass corresponding to static part of the
gravitational field of Kerr-Newman black hole.

Second part of $M$, $M_{r}= \frac {1}{2}\frac {{\it a}^{2}}{R}$,
represents classically the mass, i.e. rotation kinetic energy
corresponding to angular momentum $J={\it a}M$ and radius $R$.

We can introduce the following
\begin {equation}
 M_{g} = M_{s} + M_{r} = \frac {R}{2}+ \frac {1}{2}\frac {{\it a}^{2}}{R}= \frac { R^{2} + {\it a}^{2}}{2R }
\end {equation}
\begin {equation}
 R_{g} = 2 M_{g}         .
\end {equation}
Here $M_{g}$ can be considered as an effective mass corresponding
to total gravitational mass representing sum of the static and
rotation mass, while $R_{g}$ can be considered as horizon radius
of a fictitious Schwarzschild black hole with mass $M_{g}$.

Third part of $M$ (2), $\frac {1}{2}\frac {Q^{2}}{R}$, can be
considered as an effective mass, i.e. potential energy of the
electrostatic repulsion of the homogeneously charged thin shell
with electrical charge $Q$ and radius $R$.

Finally, we can define
\begin {equation}
  M_{red}= (M^{2} - {\it a}^{2}- Q^{2})^{\frac {1}{2}}= M(1 - \frac {{\it a}^{2}+ Q^{2}}{M^{2}})^\frac {1}{2}
\end {equation}
which can be considered as an effective, reduced black hole mass
obtained by diminishing  of the real black hole mass $M$ by means
of, classically speaking, rotation ("centrifugal force") and
electrostatic repulsion.

Suppose now that, for "macroscopic" (with mass many time larger
than Planck mass, i.e. 1) Kerr-Newman black hole, at horizon
surface there is some small (with "microscopic" masses, i.e.
masses smaller than Planck mass, i.e. 1) quantum system. It can be
supposed that given small quantum system at black hole horizon
represents the quant of the self-interaction of black hole, or,
quant of the interaction between formally separated black hole and
its fields.

Further, for a "macroscopic" Kerr-Newman black hole, instead of
the complete dynamics of its self-interaction, only stable
(stationary) dynamical situations will be considered. Given
stability will be determined by the following condition
\begin {equation}
    m_{n} R = n \frac {1}{2\pi},  \hspace{0.5cm}   {\rm for}  \hspace{0.5cm} m_{n}\ll M
    \hspace{0.5cm}{\rm and} \hspace{0.5cm}n=1,
      2,...
\end {equation}
where $m_{n}$ for  $m_{n}\ll M$  and    $n = 1, 2,... $ represent
the mass (energy) spectrum of given small quantum system. It
corresponds to expression
\begin {equation}
      2\pi R = n \frac {1}{m_{n}} = n \lambda_{rn} \hspace{0.5cm}   {\rm for}  \hspace{0.5cm} m_{n}\ll M
    \hspace{0.5cm}{\rm and} \hspace{0.5cm}n=1,
      2,...           .
\end {equation}
were $2\pi R$ represents the circumference of Kerr-Newman black
hole outer horizon while
\begin {equation}
        \lambda_{rn}=\frac {1}{m_{n}}
\end {equation}
represents $n$-th reduced Compton wavelength of mentioned small
quantum system with mass $m_{n}$ for $n = 1, 2,...$ . Expression
(7) simply means that circumference of Kerr-Newman black hole
outer horizon holds exactly $n$ corresponding $n$-th reduced
Compton wave lengths of given small quantum system with mass
$m_{n}$ captured at black hole horizon surface, for $n = 1, 2,...$
. Obviously, it is essentially analogous to well-known Bohr's
angular momentum quantization postulate interpreted via de Broglie
relation. (Moreover, in more accurate quantum mechanical analysis
Bohr-de Broglie standing waves turn out in Schrödinger stationary
quantum states, while our reduced Compton waves turn out in
quantized small oscillations of Copeland-Lahiri circular (string)
loop [4].) However, there is a principal difference with respect
to Bohr's atomic model. Namely, in Bohr's atomic model different
quantum numbers $n = 1, 2,... $, correspond to different circular
orbits (with circumferences proportional to $n^{2} = 1^{2}, 2^{2},
…$). Here any quantum number $n = 1, 2, …$ corresponds to the same
circular orbit (with circumference $2\pi R$).

According to (6) and (1) it follows
\begin {equation}
       m_{n} = n \frac {1}{2\pi R} = n \frac {1}{2\pi (M + (M^{2} - {\it a}^{2}- Q^{2})^{\frac {1}{2}})}\equiv n m_{1}
\hspace{0.5cm}   {\rm for}  \hspace{0.5cm} m_{n}\ll M
    \hspace{0.5cm}{\rm and} \hspace{0.5cm}n=1,
      2,...
\end {equation}
where
\begin {equation}
       m_{1} = \frac {1}{2\pi R} = \frac {1}{2\pi (M + (M^{2} - {\it a}^{2}- Q^{2})^{\frac {1}{2}})}
\end {equation}
represents the ground mass of small quantum system. Obviously,
$m_{1}$ depends of $M$ so that $ m_{1}$ decreases when $M$
increases and vice versa. For a "macroscopic" black hole, i.e. for
$M \gg 1$ it follows $ m_{1}\ll 1 \ll M$.

Now, it is not hard to see that, according to (3), quotient of
$M_{g}$ and $ m_{1}$  represents well-known Bekenstein-Hawking
entropy of Kerr-Newman black hole, i.e.
\begin {equation}
       S = \frac {M_{g}}{ m_{1}}= \pi (R^{2}+{\it a}^{2})  =\frac {A}{4}          ,
\end {equation}
where, according to Bekenstein supposition, $A = 4S$ represents
the black hole surface area. Obviously, it represents an
interesting dynamical interpretation of Kerr-Newman black hole
entropy whose statistical meaning will be discussed later.

Further, according to (3)-(5), (10), define
\begin {equation}
  V = - \frac {M_{red}m_{1}}{R_{g}} = - \frac {(M^{2} - {\it a}^{2}- Q^{2})^{\frac {1}{2}}}{2\pi (R^{2} + {\it a}^{2})}
\end {equation}
that can be considered as the classical potential of the
gravitational interaction between effective black hole part with
mass $ M_{red}$ and small quantum system in the ground mass state
$ m_{1}$ at distance $ R_{g}$.

Now, it can be observed that
\begin {equation}
  T = - V = \frac {M_{red}m_{1}}{R_{g}} =  \frac {(M^{2} - {\it a}^{2}- Q^{2})^{\frac {1}{2}}}{2\pi (R^{2} + {\it a}^{2})}
\end {equation}
represents well-known Hawking temperature of Kerr-Newman black
hole. It represents an interesting dynamical interpretation of
Kerr-Newman black hole temperature.

Thus Kerr-Newman black hole entropy (11) and temperature (13) are
interpreted phenomenologically dynamically in a simple,
quasi-classical way. Also, it can be observed that for a
Schwarzschild black hole, representing an especial case of
Kerr-Newman black hole for ${\it a}=0$ and $Q=0$, it becomes
satisfied $R=2M$, $M_{s} = M_{g}= M_{red}= M$. It implies $S=\frac
{M}{m_{1}}$ and $T=- V=- \frac {Mm_{1}}{R}$ representing,
intuitively, very clear and simply, "obvious", quasi-classical
interpretation of Schwarzschild black hole entropy (as quotient of
the black hole mass and small quantum system ground mass) and
temperature ( as negative classical potential energy of the
gravitational interaction between black hole and small quantum
system in mass ground state). Vice versa, clearness and
simplicity, i.e. "obviousness", of Kerr-Newman black hole entropy
(11) and temperature (13) follow from fact that they represent
simplest generalization of previously interpreted Schwarzschild
black hole entropy and temperature.

\section {Statistical meaning of Kerr-Newman black hole entropy }

However, we shall give a deeper, statistical interpretation of
Kerr-Newman black hole entropy.

Suppose that small quantum system interacting with (macroscopic)
Kerr-Newman black hole as a thermal reservoir does a bosonic great
canonical  ensemble in thermodynamical equilibrium, with mass
spectrum $m_{n}$ for $n = 1, 2, …$ (9), temperature $T$ (13) and
chemical potential $\mu$ whose value will be determined later.

Then, as it is well-known and according to (9), statistically
averaged number of the small quantum systems with mass $m_{n}$,
$N_{n}$, for $n = 1, 2, …$, is given by expression
\begin {equation}
 N_{n} = g_{n}\frac {1}{\exp[\frac {m_{n} - \mu}{T}] -1} = g_{n}\frac {1}{\exp[\frac {nm_{1} - \mu}{T}] -1}
 \hspace{1cm}   {\rm for}  \hspace{1cm}  n=1,
      2,...
\end {equation}
where $g_{n}$  represents the degeneracy of the quantum state
corresponding to $m_{n}$  \hspace{1cm}   {\rm for}  \hspace{1cm}
n=1, 2,....

Also, as it is well-known too, partial entropy in the quantum
state corresponding to $m_{n}$  for $n = 1, 2, …$,  is given by
expression
\begin {equation}
  S_{n} = g_{n} \ln [1 + \frac {N_{n}}{g_{n}}] + N_{n} \ln [1 + \frac{ g_{n}}{ N_{n}}]
  \hspace{1cm}   {\rm for}  \hspace{1cm}  n=1,
      2,...
\end {equation}
where $g_{n}$ represents the degeneracy of the quantum state
corresponding to $m_{n}$  for $n = 1, 2, …$ .

We shall suppose
\begin {equation}
   g_{n} \simeq 1  \hspace{1cm}   {\rm for}  \hspace{1cm}       n \gg 1
\end {equation}
which, according to (14), (15) implies
\begin {equation}
     N_{n}\ll 1    \hspace{1cm}   {\rm for}  \hspace{1cm}      n \gg 1
\end {equation}
and
\begin {equation}
     S_{n}\simeq N_{n}\ll 1   \hspace{1cm}   {\rm for}  \hspace{1cm}     n \gg 1              .
\end {equation}

Also, we shall suppose
\begin {equation}
    g_{1}= N_{1}                                                     .
\end {equation}
It, according (10)-(13), implies the following value of the
chemical potential
\begin {equation}
  \mu = m_{1} - T \ln 2 = m_{1} (1 - \frac {T}{m_{1}}\ln2 ) = m_{1} (1 - \frac {1-\frac {M}{R}}{1+\frac {{\it a}^{2}}{M^{2}}}\ln2 )
\end {equation}

Intuitive explanation of the suppositions (16), (19) is very
simple. Ground mass state corresponding to m1, (energetically)
closest to (outer) horizon, maximally exposed to gravitational
influence, is maximally degenerate. Highly excited quantum states
corresponding to $m_{n}$ for $n \gg 1$, (energetically) very
distant from horizon, are not so strongly exposed to gravitational
influence and are almost non-degenerate.

It can be observed that here we have a situation in some degree
similar to Bose condensation. Small quantum systems occupy
maximally, maximally degenerate ground mass state, in respect to
other, practically non-degenerate, mass states even if, according
to (19), $\frac {N_{1}}{g_{1}}$ does not tend toward infinity but
toward 1.

According to (15)-(19) it follows
\begin {equation}
  S_{1} = 2 \ln2 N_{1}\simeq 1.39 N_{1}\sim ~ N_{1}  \hspace{1cm}   {\rm for}  \hspace{1cm}      n = 1               .
\end {equation}
It implies the following expression for usually statistically
defined total entropy $S$
\begin {equation}
  S = \sum_{n=1}  S_{n} \simeq S_{1} \simeq 1.39 N_{1} \sim N_{1}
\end {equation}
and equivalence of  (22) and (11) implies
\begin {equation}
  N_{1} \simeq \frac {1}{1.39}\frac {M_{g}}{m_{1}}\simeq 0.72  \frac {M_{g}}{m_{1}}\sim \frac {M_{g}}{m_{1}}         .
\end {equation}
Then, statistically averaged total number of the small quantum
systems N is given by expression
\begin {equation}
  N = \sum_{n=1}  N_{n} \simeq N_{1} \simeq \frac {1}{1.39}\frac {M_{g}}{m_{1}}  \simeq 0.72  \frac {M_{g}}{m_{1}}\sim  \frac {M_{g}}{m_{1}}           .
\end {equation}
and statistically averaged black hole gravitational mass of the
ensemble $<M_{g}>$ is given by expression
\begin {equation}
  <M_{g}> = \sum_{n=1}  N_{n} m_{n} \simeq N_{1} m_{1}  \simeq \frac {1}{1.39} M_{g}\simeq 0.72 M_{g} \sim  M_{g}
\end {equation}
which corresponds approximately to black hole gravitational mass $
M_{g}$.

In this way we founded statistically in a satisfactory
approximation all previously discussed basic thermodynamical
characteristics of Kerr-Newman black hole. In other words,
suggested statistics yields results in a satisfactory agreement
with previous thermodynamical predictions.

However, it can be observed that supposition (16) cannot be
determined by condition (19) or some other statistical or
thermodynamical expression. For this reason we shall simply
suppose the following form of mass (energy) degeneracy in the
general case
\begin {equation}
 g_{n} = (N_{1} -1) \exp[- \frac {m_{n} - m_{1}}{T}] +1 \hspace{1cm}   {\rm for}  \hspace{1cm}  n=1,
      2,...
\end {equation}
that, for $n = 1$, is equivalent to (19), while, for $n \gg 1$, is
equivalent to (16) .

\section {Rough, qualitative description of black hole radiation}

As it has been shown previously practically all small quantum
systems from the statistical ensemble occupy ground mass quantum
state. For this reason transitions (jumps) from higher into lower,
especially ground, mass quantum state cannot be primary cause of
black hole Hawking radiation in our simple model.  In this way in
our model it must be supposed that there are some additional,
subtle dynamical processes, corresponding Hawking near horizon
particle-antiparticle creation, which cause black hole radiation
and mass decrease, on the one hand. On the other hand, given
subtle dynamical processes must be presented in our simple,
approximate model only roughly, phenomenologically. It can be done
in a way very close to Parikh and Wilczek model of Hawking
radiation as tunneling [5], or, in further conceptual analogy as a
nuclear alpha decay. [5], or, in further conceptual analogy as a
nuclear alpha decay.

Suppose that one, arbitrary, of $S$ small quantum systems in
ground mass quantum state  interacts dynamically with other $S-1$
small quantum systems in ground mass quantum state similarly as
one alpha particle with other alpha particles in the alpha
radioactive atomic nucleus. Then, like to the model of alpha decay
as quantum tunneling, given interaction can be presented as
propagation of one small quantum system in the potential barrier
(determined by black hole and other small quantum systems)
including possibility of the tunneling, i.e. small quantum system
decay.

Suppose that given individual decay occurs statistically during
some time interval $\Delta t_{1}$ and that energy of decayed small
quantum system turns out in black hole radiation. Then, according
to Heisenberg energy-time uncertainty relation it follows
\begin {equation}
   \Delta t_{1}\simeq \frac {1}{\Delta m_{1}}
\end {equation}
where $\Delta m_{1}$ represents uncertainty of the mass in the
ground mass quantum state corresponding to small quantum system
mass $m_{1}$.

Suppose that ground mass level is sharply defined, i.e. that
\begin {equation}
    \Delta m_{1} \ll  m_{1}
\end {equation}
or, for example,
\begin {equation}
   \Delta m_{1}  = \frac {1}{100} m_{1}        .
\end {equation}

Now, total time interval for black hole complete evaporation can
be roughly presented by expression
\begin {equation}
   \Delta t_{tot}\simeq S \Delta t_{1}          .
\end {equation}

In the simplest case, i.e. for Schwarzschild black hole as an
especial limit of Kerr-Newman black hole, (30), as it is not hard
to see according to or, according to (10), (11), (27), (29),
\begin {equation}
   \Delta t_{tot}\simeq 1600 \pi^{2}M^{3} = 5027 \pi M^{3}                     .
\end {equation}
It is very close to Hawking time interval for total evaporation of
the black hole
\begin {equation}
   \Delta t_{tot}= 5120 \pi M^{3}                     .
\end {equation}

In this way we demonstrated that our model, very close to Parikh
and Wilczek modeling of Hawking radiation as tunneling, is able to
describe roughly (qualitatively) and phenomenologically, but
non-trivially, black hole radiation and evaporation too.

\section {Conclusion}

In conclusion we shall repeat and point out the following.  In
this work we presented a simple, approximate method for analysis
of the basic dynamical and thermodynamical characteristics
(Bekenstein-Hawking entropy, and Hawking temperature) of
Kerr-Newman black hole. Instead of the complete dynamics of the
black hole self-interaction we considered only such stable
(stationary) dynamical situations determined by condition that
black hole (outer) horizon circumference holds the integer number
of the reduced Compton wave lengths corresponding to mass spectrum
of a small quantum system (representing quant of the black hole
self-interaction). (Obviously it is conceptually analogous to Bohr
quantization postulate interpreted by de Broglie relation in Old,
Bohr-Sommerfeld, quantum theory. Also, it can be pointed out that
our formalism is not theoretically dubious, since, at it is not
hard to see, it can represent an extreme simplification of a more
accurate, e.g. Copeland-Lahiri , string formalism for the black
hole description.) Then, we showed that Kerr-Newman black hole
entropy represents the quotient of the sum of Schwarzschild part
and rotation part of mass of black hole on the one hand and ground
mass of small quantum system on the other hand. Also we showed
that Kerr-Newman black hole temperature represents the negative
value of the classical potential energy of gravitational
interaction between part of black hole with reduced mass and small
quantum system in the ground mass quantum state. Finally, we
suggested a bosonic great canonical distribution of the
statistical ensemble of given small quantum systems in the
thermodynamical equilibrium with (macroscopic) Kerr-Newman black
hole as thermal reservoir. Also, we suggest that, practically,
only ground mass quantum state is significantly degenerate while
all other, excited mass quantum states are non-degenerate.
Kerr-Newman black hole entropy is practically equivalent to the
ground mass quantum state degeneration. Given statistical
distribution admits a rough (qualitative) but simple modeling of
Hawking radiation of the black hole too. In many aspects this
modeling is very close to Parikh and Wilczek modeling of Hawking
radiation as tunneling.

\section {References}

\begin {itemize}

\item [[1]] V. Pankovic, M. Predojevic, P. Grujic, {\it A Bohr's Semiclassical Model of the Black Hole Thermodynamics}, gr-qc/0709.1812
\item [[2]] V. Pankovic, S. Ciganovic, R. Glavatovic, {\it The Simplest Determination of the Thermodynamical Characteristics of Kerr-Newman Black Hole}, gr-qc/0804.2327
\item [[3]] V. Pankovic, S. Ciganovic,  J. Ivanovic, {\it A Simple Determination of the (Logarithmic) Corrections of Black Hole Entropy  "without Knowing the Details of Quantum Gravity"} gr-qc/0810.0916
\item [[4]] E. J. Copeland, A.Lahiri, Class. Quant. Grav. , {\bf 12} (1995) L113 ; gr-qc/9508031
\item [[5]] M. K. Parikh, F. Wilczek, {\it Hawking radiation as Tunelling}, gr-qc/9907001
\item [[6]] R. M. Wald, {\it General Relativity} (Chicago University Press, Chicago, 1984)

\end {itemize}

\end {document}